\documentclass[12pt]{article}
\usepackage[T1]{fontenc}
\usepackage{amsmath}
\usepackage{amsfonts}
\usepackage{setspace}
\usepackage{amssymb}
\usepackage{graphicx}
\usepackage{centernot}
\usepackage[round]{natbib}

\usepackage[utf8]{inputenc}
\usepackage[english]{babel}
\usepackage{hyperref}



\begin{document}
\title{\vspace{-1cm}Algorithmic Data Analytics, Small Data Matters and Correlation versus Causation\thanks{Invited contribution to the festschrift \textit{Predictability in the world: philosophy and science in the complex world of Big Data} edited by J. Wernecke on the occasion of the retirement of Prof. Dr. Klaus Mainzer, Springer Verlag. Chapter based on an invited talk delivered to UNAM-CEIICH via videoconference from The University of Sheffield in the U.K. for the Alan Turing colloquium ``From computers to life'' \url{http://www.complexitycalculator.com/TuringUNAM.pdf}) in June, 2012.}}
\author{Hector Zenil\\ Unit of Computational Medicine, SciLifeLab, Department of\\Medicine Solna, Karolinska Institute, Stockholm, Sweden;\\ Department of Computer Science, University of Oxford, UK; and\\ Algorithmic Nature Group, LABORES, Paris, France.}
\date{}

\maketitle

\begin{abstract}
This is a review of aspects of the theory of algorithmic information that may contribute to a framework for formulating questions related to complex, highly unpredictable systems. We start by contrasting Shannon entropy and Kolmogorov-Chaitin complexity, which epitomise correlation and causation respectively, and then surveying classical results from algorithmic complexity and algorithmic probability, highlighting their deep connection to the study of automata frequency distributions. We end by showing that though long-range algorithmic prediction models for economic and biological systems may require infinite computation, locally approximated short-range estimations are possible, thereby demonstrating how small data can deliver important insights into important features of complex ``Big Data''.\\

\end{abstract}

\section{Introduction}

\begin{figure}[ht!]
\centering
\includegraphics[scale=.25]{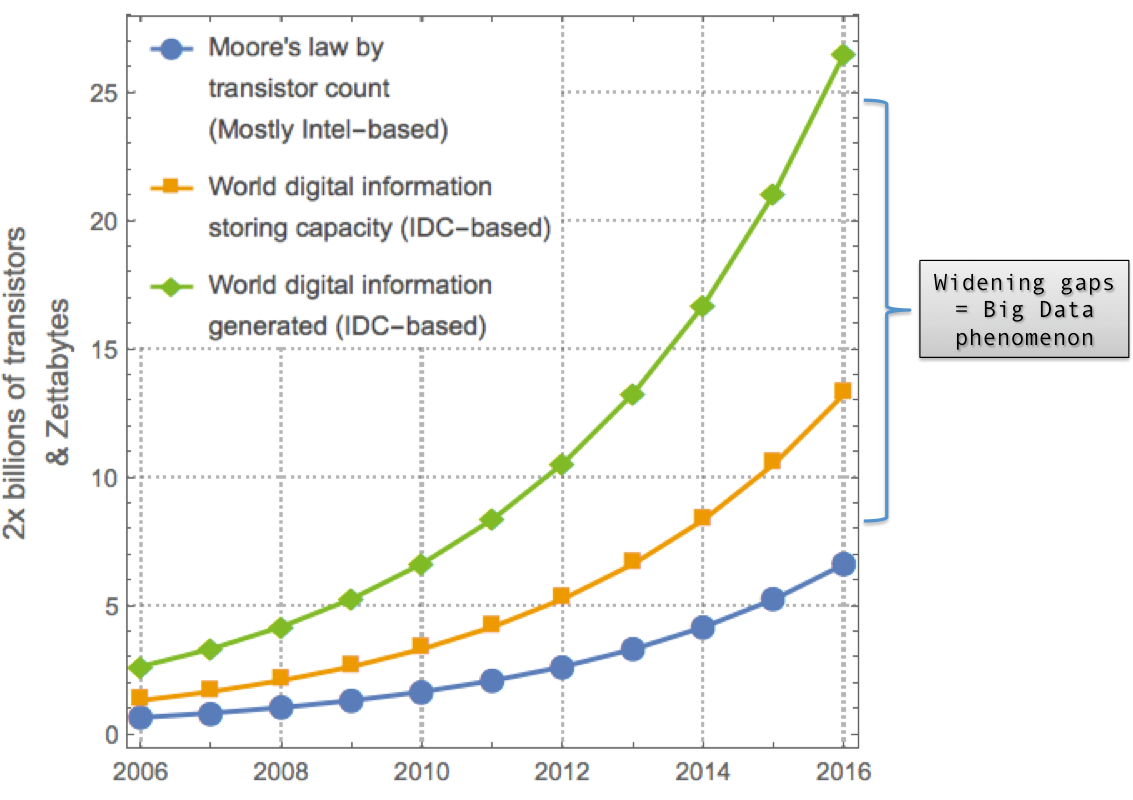}
\caption{\label{bigdata}Computational power versus generation and storage of digital data. The gap explains the `Big Data' phenomenon and the plot shows that the gap is unlikely, if not impossible, to close and will rather widen in the future. The plot can only be fair if Moore's law is also multiplied by the world number of computers but their number per capita will never grow exponentially in average by number of CPUs or GPUs in the future and so this is a constant with no real effect. Most generated digital data is algorithmic even if unstructured, that means it is the result of a causal recursive (algorithmic) mechanism, e.g. genome sequencing or TV broadcasting. IDC source: IDC's Digital Universe in 2020, sponsored by EMC, Dec 2012.}
\end{figure}

Complex systems have been studied for some time now, but it was not until recently that sampling complex systems became possible, systems ranging from computational methods such as high-throughput biology, to systems with sufficient storage capacity to store and analyse market transactions. The advent of \textit{Big Data} is therefore the result of the availability of computing power to process these complex systems, from social to economic to physical and biological systems. Much has been said on the importance of being able to deal with large amounts of data, amounts that require division into smaller segments if said data is to be analysed, and once understood, then exploited. As it can be seen in~\ref{bigdata}, generated digital information (mostly unstructured) is currently growing by 2 times the information stored but both stored and generated are outperforming computer power. 

For example, common mathematical models of the dynamics of market prices (e.g., the Black-Scholes model) assume a \emph{geometric Brownian motion}. In and of itself  (the model can and is commonly tweaked) a geometric Brownian motion implies that price changes accumulate in (log-normal) Gaussian distributions prescribing constant volatility and a controlled, risk-free financial environment. These models work fine \emph{on the average day} but are of limited use in turbulent times. In the long term, volatility is far from constant when price movement data is plotted; it is extreme price changes that bring on very rough behaviour~\citep{mandelbrot}. As shown in~\citep{zenileco}, the long-tail distribution observed in market prices may be due to behavioural decisions that resonate with algorithmic mechanisms, reflecting an algorithmic signature. \footnote{Furthermore, today more than 70\% of trading operations in the stock market are conducted automatically, with no human intervention, cf. algorithmic trading and high-frequency trading (before human traders are even capable of processing the information they register)}

The concepts and methods reviewed here are useful in cases where the complex systems' data to be divided is incomplete or noisy and cannot be analysed in full in the search for local regularities of interest whose frequency is an indication of an exploitable algorithmic signature, allowing the profiling of otherwise unmanageable amounts of data in the face of spurious correlations. This is a foundation of what I call \textit{algorithmic data analytics}.

\subsection{Descriptive complexity measures}

Central to information theory is the concept of Shannon's information entropy, which quantifies the average number of bits needed to store or communicate a message. Shannon's entropy determines that one cannot store (and therefore communicate) a symbol with $n$ different symbols in less than $\log(n)$ bits. In this sense, Shannon's entropy determines a lower limit below which no message can be further compressed, not even in principle. Another application (or interpretation) of Shannon's information theory is as a measure for quantifying the \emph{uncertainty} involved in predicting the value of a random variable.

For an ensemble $X(R, p(x_i))$, where $R$ is the set of possible outcomes (the random variable), $n=|R|$ and $p(x_i)$ is the probability of an outcome in $R$. The Shannon information content or entropy of $X$ is then given by

$$H(X)=-\sum_{i=1}^n p(x_i) \log_2 p(x_i)$$

Which implies that to calculate $H(X)$ one has to know or assume the mass distribution probability of ensemble $X$. 

The Shannon entropy of a string such as 0101010101... at the level of single bits, is maximal, as there are the same number of 1s and 0s, but the string is clearly regular when two-bit blocks are taken as basic units, in which instance the string has minimal complexity because it contains only 1 symbol (01) from among the 4 possible ones (00,01,10,11). One way to overcome this problem is by taking into consideration all possible ``granularities'' (we call this \textit{entropy rate}), from length 1 to $n$, where $n$ is the length of the sequence. To proceed by means of entropy rate is computationally expensive as compared to proceeding in a linear fashion for fixed $n$, as it entails producing all possible overlapping ${{i}\choose{n}}$ substrings for all $i \in \{1,\ldots,n\}$.

Most, if not all, classical methods and tools in data analytics are special cases, or have their roots in a version of this type of statistical  function, yet as we will see later on, it is highly prone to produce false positives of data features, as the result of spurious correlations.

\section{Computability theory for data analytics} 

A function $f$ is said not to be computable (uncomputable or undecidable) if there is no computer program running on a \emph{universal} Turing machine that is guaranteed to produce an output for its inputs or, in other words, if the machine computing $f$ doesn't halt for a number of inputs. A universal Turing machine is a general-purpose machine that can behave like any specific purpose Turing machine. The theory of algorithmic information proves that no computable measure of complexity can test for all (Turing) computable regularity in a dataset. That is, there is no test that can be implemented as a Turing machine that takes the data as input and proceeds to retrieve a regularity that has been spotted (e.g., that every 5th place is occupied by a consecutive prime number). A computable regularity is a regularity for which a test can be set as a computer program running on a specific-purpose Turing machine testing for that regularity. Common statistical tests, for example, are computable because they are meant to be effective (on purpose), but no computable \emph{universal} measure of complexity can test for every computable regularity. Hence only noncomputable measures of complexity are up to the task of detecting any possible (computable) regularity in a dataset, given enough computing time. Therefore:

\begin{center}
Universality $\Longrightarrow$ Uncomputability
\end{center}

Which is to say that universality entails or implies uncomputability, for if a system is computable (or decidable), then it cannot be universal. One may wonder why more power is needed when dealing with finite strings in limited time (real data). Though finite in length, the number of possible finite strings is infinite, and detecting only a fraction of regularities leaves an infinite number of them undetected. This is therefore a real concern at the limits of what is and is not computable. 

\subsection{Finding the generating mechanism}

The aim of data analytics (and for that matter of science itself) is to find the mechanisms that underlie phenomena, for purposes such as prediction. The optimal way to do this is to find the generating program, if any, of a piece of data that is of interest. To this end, the concept of algorithmic complexity (also known as Kolmogorov or Kolmogorov-Chaitin complexity) is key. The algorithmic complexity $K(x)$ of a piece of data $x$ is the length of the shortest program $p$ that produces $x$ running on a universal Turing machine $U$ (or a Turing-complete language, i.e., a language expressive enough to express any \textit{computable object}). Formally,

\begin{equation}
K_U(x) = min\{|p|, U(p)=x\}
\end{equation}

Data with low algorithmic complexity is compressible, while random ``data'' is not~\citep{kolmogorov,chaitin}. For example, strings like $1111111111\ldots$ or $010101010\ldots$ have low algorithmic complexity because they can be described as $n$ times 1 or $m$ times 01. Regardless of their size the description only grows by about $log(k)$ (the integer size of $n$ or $m$). The string 0110101100110110111001, however, has greater algorithmic complexity because it does not seem to allow a (much) shorter description than one that includes every bit of the string itself, so a shorter description may not exist. For example, the repetitive string $01010101\ldots$ can be produced \textit{recursively} by the following program $p$:

\begin{quotation}
\noindent \textsc{1: n:= 0}\\
\textsc{2: Print n}\\
\textsc{3: n:= n+1 mod 2}\\
\textsc{4: Goto 2}
\end{quotation}

The length of this $p$ (in bits) is thus an upper bound of the complexity $K$ of $010101\ldots01$ (plus the size of the line to terminate the program at the desired string length). Producing a shorter version of a string is a test for non-randomness, but the lack of a short description (program) is not a guarantee of randomness.

As we will discuss, $K$ is extremely powerful, if we bear in mind the caveat that it can only be approximated. Before explaining the other side of the coin, namely algorithmic probability (AP), let's study some properties of algorithmic complexity that are relevant (because AP encompasses all of them too) to an algorithmic approach to data analytics.

\subsection{Uncomputability and the choice of language}

Tailored measures of ``complexity'' have often been used for pattern detection in areas such as data analytics. Many of these ``complexity'' measures are based on Shannon's entropy. The chief advantage of these measures is that they are computable, but they are also very limited, and we are forced to focus on a small subset of properties that such a measure can test. A computable measure is implemented by an effective algorithm that, given an input, retrieves the expected output, for example, the Shannon entropy of a string. Attempts to apply more powerful \emph{universal}  (universal in the sense of being able to find any pattern in the data) measures of complexity traditionally face two challenges:

\begin{itemize}
\item Uncomputability, and
\item Instability.
\end{itemize}

On the one hand, $K$ would only make sense as a complexity measure for describing data if it showed signs of stability in the face of changes in the language used to describe an object. An \textit{invariance theorem} guarantees that in the long term different algorithmic complexity evaluations (with different description languages) will converge to the same values as the size of data grows.

If $U_1$ and $U_2$ are two (universal) Turing machines and $K_{U_1}(x)$ and $K_{U_2}(x)$ the algorithmic complexity of a piece of data $x$ when $U_1$ or $U_2$ are used respectively, there exists a constant $c$ such that for all ``binarisable'' (discrete) data $x$~\citep{kolmogorov,chaitin,solomonoff}:

\begin{equation}
| K_{U_1}(x) - K_{U_2}(x) | < c_{_{U_1,U_2}}
\end{equation}

One can think of $c_{_{U_1,U_2}}$ as the length of the computer program that translates the programs producing $x$ on $U_2$ in terms of $U_1$ and the other way around, usually called a \textit{compiler}.

In other words, evaluations of $K$ using two different descriptive languages will differ by a constant $c$.

On the other hand, uncomputability has for a long time been seen as the main ``drawback'' of $K$ (and related measures). No algorithm can tell whether a program $p$ generating $x$ is the shortest (due to the undecidability of the halting problem of Turing machines). But the uncomputability of $K(x)$ is also the source of its power, namely, the universality of this measure as based on a reference universal Turing machine. This is because measures based on non Turing-universal machines (eg finite automata) do not allow a general universal measure capable of characterizing every possible regularity in the data, including non-statistical ones, such as recursive or algorithmic (those that can compress the data by using generating mechanisms even if not obviously statistically present). While for a single piece of data only a finite number of possible patterns can be associated to that piece of data, upper bounded by its Shannon entropy (all possible combinations of the data symbols), the number of all possible patterns (statistical and algorithmic) in \textit{any} piece of data (eg strings) is uncountably infinite by a Cantor diagonalization (cf.~\ref{diag}). Martin-L{\"o}f proves~\citep{martinlof} that no computable measure can capture the typicality of all possible strings and therefore computable measures will miss specific important properties of these strings. He also proves there is a \textit{universal test} that depends on the universality of the concept of effective method implemented by a universal Turing machine.

However, to be more specific, $K$ is an upper semi-computable function, because it can be approximated \emph{from above}. Lossless compression algorithms have traditionally been used to find short programs for all sorts of data structures, from strings to images and music files. This is because lossless compression algorithms are required to reproduce the original object, while the compressed version of the object together with the decompression program constitutes a program that when decompressed reconstructs the original data. The approximation to $K$ is thus the size in bits of the compressed object together with the size in bits of the decompression program. By this means, one can, for example, find a short program for a piece of data $x$, shorter than the length $|x|$ of $x$. But even though one cannot declare an object $x$ not to have a shorter program, one can declare $x$ not random if a program generating $x$ is (significantly) shorter than the length of $x$, constituting an upper bound on $K$ (hence approximated from \emph{above}).

\subsection{Data from a system's unpredictability}

Formalised by Peter Schnorr, the unpredictability approach to randomness involves actors and concepts familiar to economists and risk managers. It establishes that a gambler cannot make money following a computable betting system $M$ against the digits of a random sequence $x$. In equational form,

\begin{equation}
\lim_{n\rightarrow \infty} M(s\uparrow n)=\infty
\end{equation}

Equivalently, a piece of data $x$ is statistically atypical (also known as Martin-L\"of random) if and only if there does not exist a computable enumerable martingale $M$ that succeeds on $x$, that is, there exists no effective computable betting strategy. It then establishes a formal connection between concepts that were previously thought to be only intuitively related:

\begin{center}
simple $\iff$ predictable\\
random $\iff$ unpredictable
\end{center}

That is, what is random is unpredictable and what is unpredictable is random; otherwise it is simple. For example, program \textsc{A} (cf. \emph{Example of an evaluation of $K$}) trivially allows a shortcut to the value of an arbitrary digit through the following function $f(n)$ (if $n=2m$ then $f(n)=1$, $f(n)=0$ otherwise)~\citep{schnorr}.

A series of \emph{universality results} (both in the sense of \emph{general} and in the sense of Turing \emph{universal}, the latter concept being a version of the former)~\citep{kirchherr} leads to the conclusion that the definition of random complexity is mathematically objective, avoiding some technical subtleties:

\begin{itemize}
\item Martin-L\"of proves~\citep{martinlof} that there is a \emph{universal} (but uncomputable) statistical test that captures all computably enumerable statistical tests. His definition of randomness is therefore general enough to encompass all effective tests for randomness.
\item Solomonoff~\citep{solomonoff} and Levin~\citep{levin} prove that the concept of \emph{universal search} (cf. algorithmic probability) is the optimal learning strategy with no prior language. In fact Levin's distribution is also called \emph{the Universal Distribution} (see Section~\ref{UD}).
\item Schnorr~\citep{schnorr} shows that a predictability approach based on martingales leads to another characterization of randomness, which in turn is equivalent to Martin-L\"of randomness.
\item Chaitin~\citep{chaitin2} proves that an uncompressible sequence is Martin-L\"of random and that Martin-L\"of randomness implies uncompressibility. That is, sequences that are complex in the Kolmogorov-Chaitin sense are also Martin-L\"of random.
\item The confluence of the above definitions (c.f. see explanation below).
\end{itemize}

One has to sift through the details of these measures to find their elegance and power (and sometimes also caveats and differences). It follows that the algorithmic characterizations of randomness converge. That means that the definitions are coextensive in the sense that the complex elements for one measure are also the complex elements for the others definitions, and the low complex elements are also low for the others. Furthermore, all the definitions assign exactly the same \emph{complexity} values to the same objects, hence they are equivalent to each other. In summary, for the most basic type of randomness, we have a more or less straight chain of simplified equivalences:

\begin{center}
uncompressibility $\Longleftrightarrow$ unpredictability $\Longleftrightarrow$ typicality
\end{center}

When this convergence happens in mathematics it is believed that a concept (in this case randomness) has been objectively captured.
 
This is a significant contribution to science which complex systems' models should take into account and build upon instead of designing new and ad hoc new measures of complexity, sometimes contradicting each other.

For algorithmic universal measures it does not matter whether the source is algorithmic and deterministic, it all boils down to the fact that if a phenomenon is unpredictable or uncompressible or statistically typical (there are no regularities in the data), then the model too will be unpredictable, uncompressible and statistically typical.

\subsection{More structured data has more short descriptions}

One preliminary observation is that even if one cannot tell when data is truly random, most data cannot have much shorter generating programs than themselves. For strings and programs written in binary a basic counting argument tells that most data must be random:

\begin{itemize}
\item There are exactly $2^n$ bit strings of length $n$, among all these strings
\item there are only $2^0 + 2^1 + 2^2 + \ldots + 2^{(n-c)} = 2^{(n-c+1)}-1$ bit strings of $c$ fewer bits. 
\item In fact there is one that cannot be compressed even by a single bit.
\item Therefore there are considerably fewer short programs than long programs.
\end{itemize}

Thus, one cannot pair off all $n$-length binary strings with binary programs of much shorter length (there simply aren't enough short programs to encode all longer strings). 

\subsection{Non-random data does not have disparate explanations}

Next to the uncomputability of $K$, another major objection to $K_U$ is its dependence on a universal Turing machine $U$. It may turn out that:

\begin{center}
$K_{U_1}(x) \neq K_{U_2}(x)$ when evaluated using $U_1$ and $U_2$ respectively.
\end{center}

This dependency is particularly troublesome for small data, e.g., short strings, shorter, for example, than the length of $T$, the universal Turing machine on which $K_T$ is evaluated (typically on the order of hundreds to thousands---a problem originally identified by Kolmogorov himself). As pointed out by Chaitin~\citep{thesis}:

\begin{quotation}
``The theory of algorithmic complexity is of course now widely accepted, but was initially rejected by many because of the fact that algorithmic complexity is on the one hand uncomputable and on the other hand dependent on the choice of universal Turing machine.''
\end{quotation}

The latter downside is especially restrictive in real world applications because this dependency is particularly true of small data (e.g., short strings). Short strings are common in various disciplines that need measures capable of exploiting every detectable regularity. Biology is one instance, specifically in the area of DNA and protein sequences, and relating particularly to identification---questions such as which genes map which biological functions or what shape a protein will fold into.

\subsection{Compressing small data to exploit big data}

A problem common to implementations of lossless compression algorithms trying to compress data is that the compressed version of the data is the compressed data together with the program that decompresses it (i.e., the decompression instructions). If the data is too small the decompression program will dominate its overall complexity length, making it difficult to produce values for comparisons for data profiling purposes, for which sensitivity is key  (e.g., the impact on the data of a small perturbation). If one wished to tell which of two pieces of small data are objectively more or less random and complex using a compression algorithm, one quickly finds out that there is no way to do so.

The constant involved $c_{_{U_1,U_2}}$ comes from the \textit{Invariance theorem} and it can be arbitrarily large, providing unstable evaluations of $K(x)$, particularly for small data (e.g., amounting to properties of interest or trends in big data). 


In~\citep{thesis,delahayezenil2,d5} we introduced a novel alternative to compression using the concept of algorithmic probability, which is proving to be useful for applications~\citep{numerical,zenil2d} and which we believe is relevant to complex systems where large amounts of data are generated.

\section{Algorithmic hypothesis testing}
\label{UD}

There is a measure that defines the probability of data being produced by a random program running on a universal Turing machine~\citep{solomonoff,levin}. This measure is key to understanding data in the best possible scientific way, that is, by exploring hypotheses in search of the most likely genesis of the data, so as to be able to make predictions. Formally,

\begin{equation}
m(x) = \sum_{p:U(p) = x} 1/2^{|p|}<1
\end{equation}
\noindent i.e. it is the sum over all the programs that produce $x$ running on a universal (prefix-free\footnote{The group of valid programs forms a prefix-free set (no element is a prefix of any other, a property necessary to keep $0 < m(x) < 1$). For details see~\citep{calude}.}) Turing machine $U$. The probability measure $m(x)$ is traditionally called Levin's semi-measure, Solomonoff-Levin's semi-measure or \textit{algorithmic probability} and it defines what is known as the \textit{Universal Distribution} (\emph{semi} from \emph{semi-decidable} and the fact that $m(x)$, unlike probability measures, doesn't add up to 1). $m(x)$ formalises the concept of Occam's razor (favouring simple---or shorter---hypotheses over complicated ones). algorithmic probability is deeply related to another fascinating uncomputable object that encodes an infinite amount of uncomputable and accessible information.

\subsection{Real big data and Chaitin's Omega \emph{number}}

$m$ is closely related to Chaitin's halting probability, also known as Chaitin's $\Omega$ number~\citep{chaitin} defined by:

\begin{equation}
\Omega_U = \sum_{U(p) \textnormal{ halts}} 1/2^{|p|}
\end{equation}
\noindent (the halting probability of a universal \emph{prefix-free} Turing machine $U$)

Evidently, $m_U(x)$ provides an approximation of $\Omega_U$ plus the strings produced (like $\Omega_U$, $m(x)$ is uncomputable, but just as is the case with $\Omega_U$, $m(x)$ can sometimes be calculated up to certain \emph{small} values).

While a Borel normal number locally contains all possible patterns, including arbitrarily long sequences of 0s or 1s, or any other digit, thereby containing all possible statistical correlations, all of them will be spurious because a Borel normal number has no true global statistical features (given the definition of Borel normality). A Chaitin $\Omega$, on the other hand, epitomises \textit{algorithmic Big Data}, where no statistical correlation is an algorithmic feature of the data. This is because a Chaitin $\Omega$ number is not only Borel normal but also algorithmically random, and therefore cannot be compressed. In fact, there are Chaitin $\Omega$ numbers for which none of the digits can be estimated and are therefore perfectly algorithmically random. Algorithmic Big Data is a subset of statistical Big Data and therefore reduces the likelihood of spurious hypotheses based on correlation results. The move towards algorithmic tools is therefore key to the practice of data science in the age of Big Data and complex systems.

\subsection{Optimal data analytics and prediction}

The concept of algorithmic complexity addresses the question of the randomness content of individual objects (independent of distributions). Connected to algorithmic complexity is the concept of algorithmic probability (AP), that addresses the challenge of hypothesis testing in the absence of full data and the problem of finding the most likely process generating the data (explanation). algorithmic probability and algorithmic complexity are two faces of the same coin.

While the theory is very powerful but is hard to calculate estimations of complexity, it does make two strong strong assumptions that are pervasive in science and that does not mean they are necessarily right but seemingly reasonable (any other assumption seems less reasonable):

\begin{itemize}
\item The generating mechanism is algorithmic as opposed to uncomputable, e.g., the result of an `oracle' (in the sense of Turing, i.e. a non-computable source), magic, or divine intervention.
\item The likelihood distribution of programs is a function of their length (formalisation of Occam's razor)
\end{itemize}

The assumptions are reasonable because consonant with the original purpose of analysing the data with a view to identifying the generating mechanisms that would explain them, and while in assigning greater probability to shortest explanations consistent with the data (Occam's razor) they do favour the simplest computer programs, they also entail a Bayesian prior for predicting all the generating programs (all possible explanations accounting for the data regularities), hence complying with Epicurus' principle of multiple explanations. While one can frame criticisms of these principles, both principles are common and among the most highly regarded principles in science, thus transcending the practicalities of performing algorithmic data analytics. One of the chief advantages of algorithmic probability as an implementation of Occam's razor is that it is not prone to over-fitting. In other words, if the data has spurious regularities, only these and nothing else will be captured by an algorithmic probability estimation, i.e., by definition no additional explanations will be wasted on the description of the data.

\subsection{Complexity and frequency}

The intuition behind the Coding theorem (~\ref{bdm}) is a beautiful relation between program-size complexity and frequency of production. If you wished to produce the digits of a mathematical constant like $\pi$ by throwing digits at random, you would have to produce every digit of its infinite irrational decimal expansion. If you seated a monkey at a typewriter (with, say, 50 keys), the probability of the monkey typing an initial segment of 2400 digits of $\pi$ by chance is $(1/50^{2400})$. 

If, instead, the monkey is seated at a computer, the chances of its producing a program generating the digits of $\pi$ are on the order of $1/50^{158}$, because it would take the monkey only 158 characters to produce the first 2400 digits of $\pi$ using, for example, C language.

So the probability of producing $x$ or $U$ by chance so that $U(p)=x$ is very different among all (uniformly distributed) data of the same length is $1/2^{|x|}$, and the probability of finding a binary program $p$ producing $x$ (upon halting) among binary programs running on a Turing machine $U$ is $1/2^{|p|}$ (we know that such a program exists because $U$ is a universal Turing machine).

The less random a piece of data the more likely it is to be produced by a short program. There is a greater probability of producing the program that produces the data, especially if the program producing $x$ is short. Therefore, if an object is compressible, then the chances of producing the compressed program that produces said object are greater than they would be for a random object, i.e., $|p| \ll |x|$ such that $U(x)=x$. 
The greatest contributor to the sum of programs $\Sigma_{U(p)=x} 1/2^{|p|}$ is the shortest program $p$, given that this is when the denominator $2^{|p|}$ reaches its smallest value and therefore $1/2^{|p|}$ reaches its greatest value. The shortest program $p$ producing $x$ is none other than $K(x)$, the algorithmic complexity of $x$. 

For $m(x)$ to be a probability measure, the universal Turing machine $U$ has to be a prefix-free Turing machine, that is, a machine that does not accept as a valid program one that has another valid program as its beginning. For example, program 2 starts with program 1, so if program 1 is a valid program then program 2 cannot be valid. 

The set of valid programs is said to constitute a prefix-free set, that is, no element is a prefix of any other, a property necessary to keep $0 < m(x) < 1$. For more details see the discussion of Kraft's inequality, in~\citep{calude}.

\subsection{Massive computation in exchange for an arbitrary choice}
\label{ctm}

In order to truly approximate Kolmogorov-Chaitin complexity rather than, for example, entropy rate (which is what is achieved by currently popular lossless compression algorithms based on statistical regularities---repetitions---captured by a sliding window traversing a string), we devised a method based on algorithmic probability~\citep{delahayezenil2,d5}.

Let $(n,2)$ be the set of all Turing machines with $n$ states and 2 symbols, and let $D(n)$ be the function that assigns to every finite binary string $x$ the quotient:

\begin{equation}
\frac{\text{\# of times that a machine $(n,2)$ produces $x$}}{\text{\# of machines in $(n,2)$}}
\end{equation}

As defined in~\citep{delahayezenil2,d5}, $D(n)$ is the probability distribution of the data (strings) produced by all $n$-state 2-symbol Turing machines. Inspired by $m$, $D(n)$ is a finite approximation to the question of the algorithmic probability of a piece of data to be produced by a random Turing machine up to size $n$. Like $m(x)$, $D(n)$ is uncomputable (by reduction to Rado's Busy Beaver problem), as proven in~\citep{delahayezenil2}. Examples for $n=1,n=2$ (normalised by the \# of machines that halt):

\begin{center}
$D(1) = 0 \rightarrow 0.5; 1 \rightarrow 0.5$\\
$D(2) = 0 \rightarrow 0.328; 1 \rightarrow 0.328; 00 \rightarrow .0834 \ldots$\\
\end{center}

By using the Coding theorem Eq.~\ref{coding} one can evaluate $K(x)$ through $m(x)$, which reintroduces an additive constant. One may not get rid of the constant, but the choices related to $m(x)$ are less arbitrary than picking a universal Turing machine directly for $K(x)$, and we have proven that the procedure is not only theoretically sound but stable~\citep{zenil2d}, and in accordance with strict program-size complexity and compressibility~\citep{numerical}, the other traditional method for approximating $K(x)$, which fails for small data that an approximation of $m(x)$ handles well. The trade-off, however, is that approximations of $m(x)$ require extraordinary computational power. Yet we were able to calculate relatively large sets of Turing machines to produce $D(4)$ and $D(5)$~\citep{delahayezenil2,d5}. $D$ can be seen as averaging $K$ over a large set of possible languages in order to reduce the possible impact of the constant from the Invariance theorem. We call CTM to the calculation of $D$ and stand for the \textit{Coding theorem method}.

It is common that an approach to deal with the involved constant in the Invariance theorem is to choose a small \textit{reference} universal Turing machine on which all evaluations of $K$ are made, but an arbitrary choice is still made. There is another approach to this challenge, and that is to exchange the choice of a reference machine for what may appear a less arbitrary choice of enumerating scheme. One can take the quasi-lexicographical enumeration as a natural choice because it sorts all possible Turing machines by size, hence consistent with the theory itself. Moreover, one can run all Turing machines of certain size at the same time and get a perfect sample of the whole space. The choice is still arbitrary in the sense that one could enumerate all possible Turing machines (which is itself a Turing universal procedure) in any other way, but the choice is \textit{averaged} over a large number of small Turing machines.

\section{Correlation versus causation}
\label{diag}

On the one hand, it has recently been shown \citep{calude2} that for increasing amounts of data, spurious correlations will increase in number and distinguishing between false and true positives becomes more challenging. On the other hand, while Shannon entropy is a very common and, in some applications, a powerful tool, it can be described as a counting function that involves a logarithm. As such, it can count arbitrary elements. In a network, for example, it can reckon edge density, edge count, connected edge count, etc., from which it is clear that not all definitions converge, i.e., given one definition it is not always true that high or low entropy will remain high or low entropy for some other definition.

According to Shannon entropy, specifying the outcome of a fair coin flip (two equally likely outcomes) requires one bit at a time, because the results are \textit{independent} and therefore each result conveys maximum entropy. This \textit{independence} is generally in appearance only, and has to be construed as \textit{statistical independence}.

While both definitions, entropy rate and Kolmogorov-Chaitin (algorithmic) complexity, are asymptotic in nature, they are essentially different, and connected in a rather trivial fashion. Low Shannon entropy implies an exploitable regularity that can be captured by recursion, i.e., a small computer program. However, maximal entropy rate (\emph{Shannon randomness}) does not imply algorithmic randomness.

By means of the so-called \textit{universal statistical test} (which tests for all effective---computable---statistical tests using a universal Turing machine), Martin-L\"of proved that algorithmic randomness can test for every effective (computable) property of an infinite sequence. In practice, one does not apply such a test to infinite objects. However, the number of finite objects is classically numerable as infinite, thus leading to an infinite number of different features. In the context of networks, for example, every network can be a feature by itself, testing membership to a set with only itself in it, but in general one may be interested in properties such as clustering coefficients, motifs, geodesics, certain kinds of betweenness, etc., an infinite number of properties. This has significant implications for data science and big data.

We say that a function $\Delta$ is robust if for an object $s$, $\Delta_T$ is invariant to the description $T$, where $T$ is a description that fully recovers $s$. In other words,\\

\noindent \textbf{Definition:} Let $\Delta_T(s)$ from $\Delta_S(s)$ be the same function applied to the same object $s$ described using descriptions T and S. If $|\Delta_T - \Delta_S| > c$ then $\Delta$ is not a robust measure. Which is to say that a robust definition is a definition that is at most a constant away from each other, and thus grows in the same direction and rate for the same objects.\\

This definition is a generalization of the \textit{Invariance theorem} in the theory of algorithmic information that does not require $T$ or $S$ to be computer programs to be run on universal Turing machines and therefore allow $\Delta$ to be also a computable function (i.e. one for which there is always a value for every evaluation of $\Delta$).

By the \textit{Invariance theorem}, Kolmogorov-Chaitin complexity and algorithmic probability are robust as they are invariant up to a constant to different object descriptions.\\

\noindent \textbf{Observation 1:} Entropy rate is more robust than Shannon entropy for an arbitrary choice of domain (tuple or $n$-gram size), but entropy rate is not invariant to description changes (e.g., 1-grams versus n-grams, or edges vs connected edges).\\

\noindent \textbf{Observation 2:} According to Martin-L{\"o}f's results, for every effective (computable) property $X$, there is a computable measure $Y$ to test whether every system $x$ has a property $X$, but there exists a property $Z$ for some $x$ that $Y$ cannot test. The results also imply that for a measure to test every effective statistical property of any possible system the measure has to be uncomputable, and that any uncomputable measure of lowest degree implemented on a universal Turing machine will only differ by a constant from any other uncomputable measure of the same type. Hence, all uncomputable complexity measures are asymptotically the same.\\

In other words, for every effective feature, one can devise/conceive an effective measure to test it, but there is no computable measure (e.g., entropy rate) able to implement the \text{universal statistical test}, only a measure of algorithmic randomness (e.g., Kolmogorov-Chaitin complexity or algorithmic probability).

That no computable function can detect all possible (enumerable effective) computable regularities can be proven by contradiction using a diagonalization argument. By definition every computable regularity can be implemented as a computable test that retrieves 0 or 1 if the regularity is present in the input. Let $q$ be an enumeration of all Turing machines, each implementing a computable test and let $r$ be an enumeration of the computable regularities. We then take the negation of every $(i,i)$ element (the diagonal) to build a (enumerable effective) computable regularity $n$ that is not in $r$. Then we arrive at the contradiction that $n \notin r$ which was supposed to include all the (enumerable effective) computable regularities. We know that all Turing machines are enumerable therefore the assumption that $r$ was enumerable has to be false. This means that no computable function can enumerate all possible regularities, not even only the (enumerable effective) computable ones. Constructive proofs are also possible by reducing the problem to the undecidability of the \emph{halting problem} (encoding the halting probability as a computable test and then arriving at a contradiction).

The power of algorithmic complexity over Shannon entropy can be expressed in the following statements of logical implication:

\begin{itemize}
\item Presence of statistical regularities $\implies$ low Shannon entropy.
\item low Shannon entropy $\implies$ low Kolmogorov-Chaitin complexity
\item Low Kolmogorov complexity $\centernot\implies$ low Shannon entropy.
\end{itemize}

To illustrate the above, let $\omega$ be the Thue-Morse sequence obtained by starting with 0 and recursively appending the Boolean complement of the sequence. The first few steps yield the initial segments 0, 01, 0110, 01101001, 0110100110010110, and so on.

Notice that, in fact, the Kolmogorov complexity of a string and its Shannon entropy are exactly the same, but for this to hold, in the calculation of Shannon entropy one needs full knowledge to the deterministic nature of the object's generating mechanism, and thus implies that in practice, in the lack of certainty, Shannon entropy will tend to over-estimate the complexity of an object.

So $\omega$ is of low Kolmogorov complexity without knowing the source (or the distribution that the source can generate) because the stream of observations can be compressed by the recurrence relation: $t_0 = 0$ or 1, $t_{2n} = t_n$, and $t_{2n+1} = 1 - t_n$ for all positive integers $n$ implemented in a program of $\log(i)$ growing size with $i$ the number of digits wished to be produced.

$\pi$ is another example of a similar type. It is believed to be Borel normal and therefore of maximal entropy like most real numbers~\citep{martinlof}, yet $\pi$ is of the lowest Kolmogorov-Chaitin complexity because small formulae can generate arbitrarily long initial segments of $\pi$. So, in the observation of the digits of $\pi$, its Shannon entropy rate is high but because any formula of $\pi$ can only generate $\pi$ and nothing else, each digit comes with no uncertainty only if, we were to know that any observed sequence of the digits of $\pi$ are actually $\pi$. 

In other words, while Shannon entropy is some sort of zero-one measure, where something apparently random can actually be deterministic, and thus with either maximal or minimal entropy (and entropy rate), the Kolmogorov complexity of observations of segments of $\pi$ would intrinsically allow the recognition of such segments as having low complexity because they have, in principle (even if hard to calculate), short computer programs that produce any of these segments (actually without having to produce any previous segments, per the BBP formulae (\cite{bailey})). 

While Shannon entropy is ill-suited (see e.g. \cite{zenilnarsis}) to intrinsically identify the causal source of an object (i.e. only by looking at the object and nothing else) beyond simple statistical patterns (repetitions) and to update the degree of belief in the (un)certainty of the generating (non)deterministic mechanism, Kolmogorov complexity does incorporate better the (un)certainty belief that it can assign to a generating process from observing its output, without having to consider the distribution of possible outputs that it is always difficult, if not impossible, to know in advance (e.g. by looking at some segments of $\pi$ to know if it is $\pi$ or not). A beautiful example illustrating these limitations of entropy is introduced in \cite{zenilnarsis}.

It is therefore clear how, unlike other computable descriptive complexity measures such as Shannon entropy, algorithmic complexity measures are not assigned maximal complexity (e.g., entropy rate), but rather low complexity if they have a generating mechanism, thereby constituting a causal origin as opposed to a random one, despite their lack of statistical regularities.

In fact, Shannon entropy can be equated with the existence of statistical regularities in an object (e.g., a sequence) relative to other elements (the elements of interest to be counted) in a distribution. Shannon entropy can also be seen as a counting function, one that can count any element or property of interest but only one at a time (modulo properties that are related or derived), relative to the occurrence of the same property in other elements in a set (the distribution).

\section{Divide-and-conquer data analysis}
\label{bdm}

The greatest contributor to the sum of programs $1/2^{|p|}$ is the shortest program $p$, and the shortest program $p$ producing $x$ is none other than $K(x)$, the algorithmic complexity of $x$. 

The algorithmic Coding theorem describes the reverse connection between $K(x)$ and $m(x)$~\citep{levin,chaitin}:

\begin{equation}
\label{coding}
K(x) = - \log_2(m(x)) + O(1)
\end{equation}

\noindent where $O(1)$ is a linear term independent of $x$. 

This tells us that if a piece of data $x$ is produced by many programs, then there is also a short program that produces $x$~\citep{cover}.

The Block decomposition method (BDM) that we introduced in ~\citep{physicaa,d5}, allows us to extend the power of the Coding theorem by using a combination of Shannon entropy and algorithmic probability. The idea is to divide a piece of data into segments for which we have algorithmic probability approximations, and then count the number of times that a local regularity occurs. We then apply the formula $\sum_{\forall x} C(x) + \log M(x)$, where $C(x)$ is the estimation of the complexity from CTM based on algorithmic probability and related to its Kolmogorov-Chaitin complexity, and $M(x)$ is the number of times such a pattern repeats. In other words, one does not count $K(x)$ $M(x)$ times but only once, because the repetitions can be produced by the addition of another program that repeats the same pattern and requires no greater description length than the shortest program producing $x$ and the logarithm of the number of repetitions. We have shown that by applying the method we can profile data in an accurate fashion~\citep{narsis2}.

\subsection{The algorithmic Bayesian framework}

We claimed that Kolmogorov complexity can update its degree of belief in the deterministic nature of the source of a generated output intrinsically from the observation of the output itself. Here we explain how a Bayesian framework based on algorithmic probability as related to Kolmogorov complexity, can provide a framework exploiting this power.

Let's come back to attempting to approximate Kolmogorov complexity by chunking data. One can use BDM to estimate the complexity of a piece of data in order to attempt to reveal its possible generating mechanism (the possible cause of the regularity). The Bayesian application is as follows. Let $x$ be the observed data, $R$ a random process and $D$ the possible generating mechanism (cause). The probability that $x$ is generated by $R$ can be estimated from Bayes' theorem:

$$P(R|x)= \frac{P(x|R)P(R)}{P(x|R)P(R)+P(x|D)P(D)},$$

\noindent where $D$ stands for ``not random'' (or deterministic) and $P(x|D)$ the likelihood of $x$ (of length $l$) being produced by a deterministic process $D$. This likelihood is trivial for a random process and amounts to $P(x|R) = 1/m^l$, where $m$, as above, is the size of the alphabet. The algorithmic approach, however, suggests that we plug in the complexity of $x$ as a normative measure (equivalently, its algorithmic probability) as a natural formal definition of $P(x|D)$. $P(x|D)$ is then the distribution we get from running randomly chosen algorithms, with (by definition) simpler strings having greater higher $P(x|D)$ on the other hand, $P(x|R)$ is the ordinary probability distribution, with equal weight on any string.

To facilitate the use of this approach, we have made available an R package called ACSS (\texttt{likelihood\_d()} returns the likelihood $P(x|D)$, given the actual length of $x$) and an Online Complexity Calculator \url{http://www.complexitycalculator.org} (the R package documentation is also in the Complexity Calculator website). This was done by taking $D(x)$---a function based on the distribution of programs produced by a large set of small random Turing machines---and normalising it with the sum of all $D(x_i)$ for all $x_i$ with the same length $l$ as $x$ (note that this also entails performing the symmetric completions).

With the likelihood at hand, we can make full use of Bayes' theorem (the R package \texttt{acss} contains the corresponding functions). One can either obtain the likelihood for a random rather than a deterministic process via function \texttt{likelihood\_ratio()} (the classical default for the prior is $P(R) = 0.5$). Or, if one is willing to make assumptions as to the prior probability with which a random rather than a deterministic process is the generating mechanism for the data $x$ (i.e., $P(R) = 1-P(D)$), one can obtain the posterior probability for a random process given $x$, $P(R|x)$, using \texttt{prob\_random()}. 

The numerical values of the complexity of $x$ are calculated by the Coding theorem method (CTM). CTM, however, is computationally very expensive (it grows faster than any computable function, equivalent to the Busy Beaver) and is ultimately uncomputable beyond small computer programs and therefore works at the short sequence granularity. We have shown, however, that CTM can profile objects by their family by simply  looking at these short sequences (small data) as local regularities, just like other approaches (e.g., network motifs~\citep{alon} shown to unveil underlying biological and physical mechanisms). Moreover, the BDM method described in Section~\ref{bdm} extends the power of CTM, its chief disadvantage being that it converges to the Shannon entropy rate the longer the sequence, if CTM (which, again, is computationally expensive but need only  be run once) is not further executed. Nevertheless, we have produced applications that are proving to be useful in data analysis and data profiling in fields ranging from cognitive science~\citep{gauvrit0,gauvrit1,gauvrit3,gauvrit4} to network theory~\citep{physicaa,narsis1,narsis2,narsis3}, to image classification~\citep{zenil2d}, molecular biology and finance~\citep{zenileco}.

\section{Further discussion and conclusion}

Data is usually not generated at random but by a process. That is why, if asked, you would say you expected a `$0$' to follow 01010101, rather than a `1' (even though, according to probability theory, both have exactly the same classical probability of occurring among all strings of the same length). 

In data generated by an algorithmic process, $m(x)$ would establish the probability of an occurrence of $x$ in the data in the presence of any other additional information. $m$ thus tells us that patterns which result from simple processes (short calculations) will be likely, while patterns produced by long uncompressible calculations are unlikely if the data is produced by a mechanistic cause (e.g., a type of customer preference for a product in a catalogue).

$m$ is not simply a probability distribution establishing that there are some patterns that have a certain probability of occurring. Unlike traditional classical distributions, it also determines the specific order in which the elements are distributed, with those of low complexity more likely to be the result of a cause and effect process. On the other hand, those that are random looking and less interesting because not the result of an underlying behaviour or pattern will appear infrequently.

For example, market prices are considered non-random, even though models may be random, because prices follow laws of supply and demand. However, since regularities are erased by the same laws, making them look random, they would nevertheless appear to be algorithmic. Market patterns are artificially and quickly erased by economic activity itself, in a natural movement towards equilibrium. Shannon entropy will tend to over-fit the amount of apparent randomness in such a social phenomenon, despite the underlying algorithmic footprints that are exploitable in principle.

In the age of complex systems and Big Data, therefore, an algorithmic account is key to coping with complex systems and spurious correlations in Big Data. We strongly suggest that these algorithmic measures constitute a viable framework for analysing data from complex systems and to move away from the traditional practice of correlation and regression so pervasive, and often damaging if not misleading, in the natural and social sciences.

\section*{Acknowledgements}

Part of this work was done while the author was visiting the Institute for Mathematical Sciences, National University of Singapore (NUS), as part of the \textit{Algorithmic Randomness} program in 2014. The visit was supported by NUS and the Foundational Questions Institute (FQXi).

\bibliography{bibliography}
\bibliographystyle{chicago}

\newpage

\subsection*{Author information}

Dr. Dr. Hector Zenil (BSc Math, UNAM; Masters Logic, Paris 1 Sorbonne; PhD Computer Science, Lille 1; PhD Philosophy of Science, Paris 1/ENS Ulm/CNRS) has published around 50 publications in indexed journals, proceedings and scholar volumes. He has held positions at the Behavioural and Evolutionary Lab at the University of Sheffield, UK (Postdoctoral Researcher); at Wolfram Research (Senior Research Associate); at the Department of Computer Science at the University of Oxford (JTF Research Fellow and Principal Investigator); and at the Unit of Computational Medicine, SciLifeLab, Center of Molecular Medicine at Karolinska Institute in Stockholm, Sweden (Assistant Professor). He has also held visiting positions at MIT, Carnegie Mellon University (CMU) and the National University of Singapore (Visiting Professor). He is also the Head of the Algorithmic Nature Group of LABoRES in Paris, France. He is the editor of \textit{A Computable Universe} (foreword by Sir Roger Penrose) and \textit{Randomness Through Computation}, both academic bestsellers according to World Scientific and Imperial College Press, among other volumes.

\end{document}